\documentclass{sdob9proc}

\newcommand{\gsim}{\raisebox{-1ex}{$\stackrel{{\displaystyle>}}{\sim}$}}

\newcommand{\msun}{${\mathrm{M}}_{\odot}$}

\title{No close companions to a sample of bright sdB stars}
\author{Roberto Silvotti,$^{1}$
        Roy\,H. \O stensen,$^{2}$
	John\,H. Telting,$^{3}$}

\affiliation{
$^{1}$ INAF-Osservatorio Astrofisico di Torino, Strada dell'Osservatorio 20, 
10025, Pino Torinese, Italy\\
$^{2}$ Department of Physics, Astronomy and Materials Science, Missouri State 
University, 901 S. National, Springfield, 
\\
\hspace{1.4mm}
MO 65897, USA\\
$^{3}$ Nordic Optical Telescope, Rambla José Ana Fernández Pérez 7, 
E-38711 Breña Baja, Spain}

\shorttitle{No close companions to a sample of bright sdB stars}
\shortauthors{Roberto Silvotti \textit{et al.}}

\abs{In this article we present preliminary results on the upper limits to 
the mass of a companion for 8 apparently single subdwarf B stars, derived from 
high-precision radial velocity measurements obtained from Harps-N spectra.
These limits, corresponding to a few Jupiter masses, show that these stars
do not have close companions and keep open the unresolved question of the 
mechanism that caused the almost complete loss of the envelope for the giant 
precursors of these stars.
For 4 stars with a larger data coverage, it was also possible to set upper
limits to the mass of a more massive companion in a wider orbit.}

\begin{document}

\maketitle

\section{Introduction}

Hot subdwarf B (sdB) stars are core-helium-burning stars with a very thin
H-rich envelope, located on the extreme horizontal branch \citep{heber16}.
Their precursors lost almost all of their envelope near the tip of the red 
giant branch and such huge mass loss is compatible with a common envelope 
ejection if the sdB precursor was in a close binary.
And indeed, of the sdB stars that are not obvious spectroscopic binaries, 
about half are found in close binaries, either with M-dwarf ore white dwarf 
companions.

However,  for the remaining apparently single sdB stars, it is not clear 
how the envelope loss happened.
This is why it is important to verify if these apparently single sdB stars 
have a low-mass companion or not.
To search for close companions to sdB stars, including substellar objects,
is the main motivation of this work.

The results presented in this article are preliminary in the sense that the
Harps-N high-resolution spectra show a large number of metal absorption lines,
many of which do not yet have a certain identification.
The line identification work is in progress and therefore the results presented
here, based on the cross correlation function (CCF) method, may undergo slight 
modifications when we will repeat our analysis using an updated set of well 
identified absorption lines.

\section{Selection of the targets}

The limiting magnitude to obtain a high S/N ratio (\gsim 50) with Harps-N is 
around 12.
When this project started a few years ago, the number of apparently single
sdB stars brighter than 12 and well visible from La Palma was around 14 and
therefore we selected all of them.
Only 9 were observed with at least 3 spectra in different epochs and one of 
them seems to be in a binary system with an M dwarf and we are waiting for 
$TESS$ data to confirm that it is a binary.
In this paper we show the results for the remaining 8 stars.



\section{Radial velocities}

For each star we collected at least 3 Harps-N spectra at different random 
epochs, with exposure times ranging from 750\,s for the brightest star 
(HD\,149382) up to 1 hour (or 4200\,s in one case) for the faintest objects.
We obtained S/N ratios at $\sim$4700 \AA\ between 30 and 140 with most values
between 60 and 120.
Target names, number of spectra, mean S/N ratios and mean radial velocities 
(RVs) are listed in Table~\ref{tab}.
We note that one star, BD+48$^\circ$2721, is a high-velocity star 
with a mean RV of --187 km/s.

\begin{table}[th]
\centering
\caption{Stars observed}
\label{tab}
\begin{tabular*}{1.0\linewidth}{l r r r}
\noalign{\smallskip}\hline\hline\noalign{\smallskip}
Name & \# of ~~& ~~~mean S/N           & mean RV \\
     & spectra &  @\,4700\,\AA~         & (m/s)~~~  \\
\noalign{\smallskip}\hline\noalign{\smallskip}
HD\,4539          & 11~~~ & 101.9\,~~~   &  --3\,392.7\,~  \\
PG\,0342+026      &  4~~~ &  71.3\,~~~   &   14\,073.0\,~  \\
UVO\,0512-08      &  3~~~ &  65.0\,~~~   &   10\,214.5\,~  \\
PG\,0909+276      &  7~~~ &  31.6\,~~~   &   18\,610.7\,~  \\
PG\,1234+253      &  3~~~ & 110.8\,~~~   &  --6\,674.0\,~  \\
HD\,149382        & 16~~~ & 106.5$^*$~~  &   23\,772.3$^*$ \\
PG\,1758+364      &  4~~~ &  64.5\,~~~	 & --31\,370.9\,~  \\
BD\,48$^\circ$2721&  3~~~ &  75.5\,~~~	 &--187\,136.7\,~  \\
\noalign{\smallskip}\hline\noalign{\smallskip}
\multicolumn{4}{l}{\small
$^*$ excluding the outlier at BJD 2457076.67, see Fig.~\ref{fig6}}
\end{tabular*}
\end{table}

The spectra were reduced and wavelength-calibrated with the Harps-N pipeline.
The pipeline includes also the possibility to compute automatically the
radial velocities using the CCF.
However, the pipeline selects the absorption lines to be used in the CCF 
computation by means of a mask corresponding to the spectral type.
And currently available masks exist only for stars of spectral class 
G2, K0, K5, M2 and M4.
Therefore we built ad hoc masks for sdB stars in order to select the good lines
to be used for the CCF computation.
Typically, we used $\approx$200 absorption lines for each star in the blue 
part of the spectrum obtaining typical RV errors of 20 to 70 m/s.
A sample CCF is shown in \citet{silvotti14} while the RV
curves are shown in the upper panels of Fig.~\ref{fig1}--\ref{fig8}.
Two stars, HD\,4539 and PG\,0342+026 (Fig.~\ref{fig1} and \ref{fig2}) show 
significant RV variations and indeed, after the Harps-N observations, both were
found to vary also photometrically from $K2$ and $TESS$ observations 
respectively, due to g-mode pulsations \citep{silvotti19, sahoo19}.

\section{Companion mass}

In order to set upper limits to the mass of a companion,
we computed a series of synthetic RV curves for different orbital periods
and companion masses, assuming circular orbits and a stellar mass of 
0.47~\msun\ (0.40~\msun\ only for HD\,4539, \citealt{silvotti19}), 
and compared these curves with the RV measurements.

\begin{figure}[p]
\includegraphics[width=1.0\linewidth]{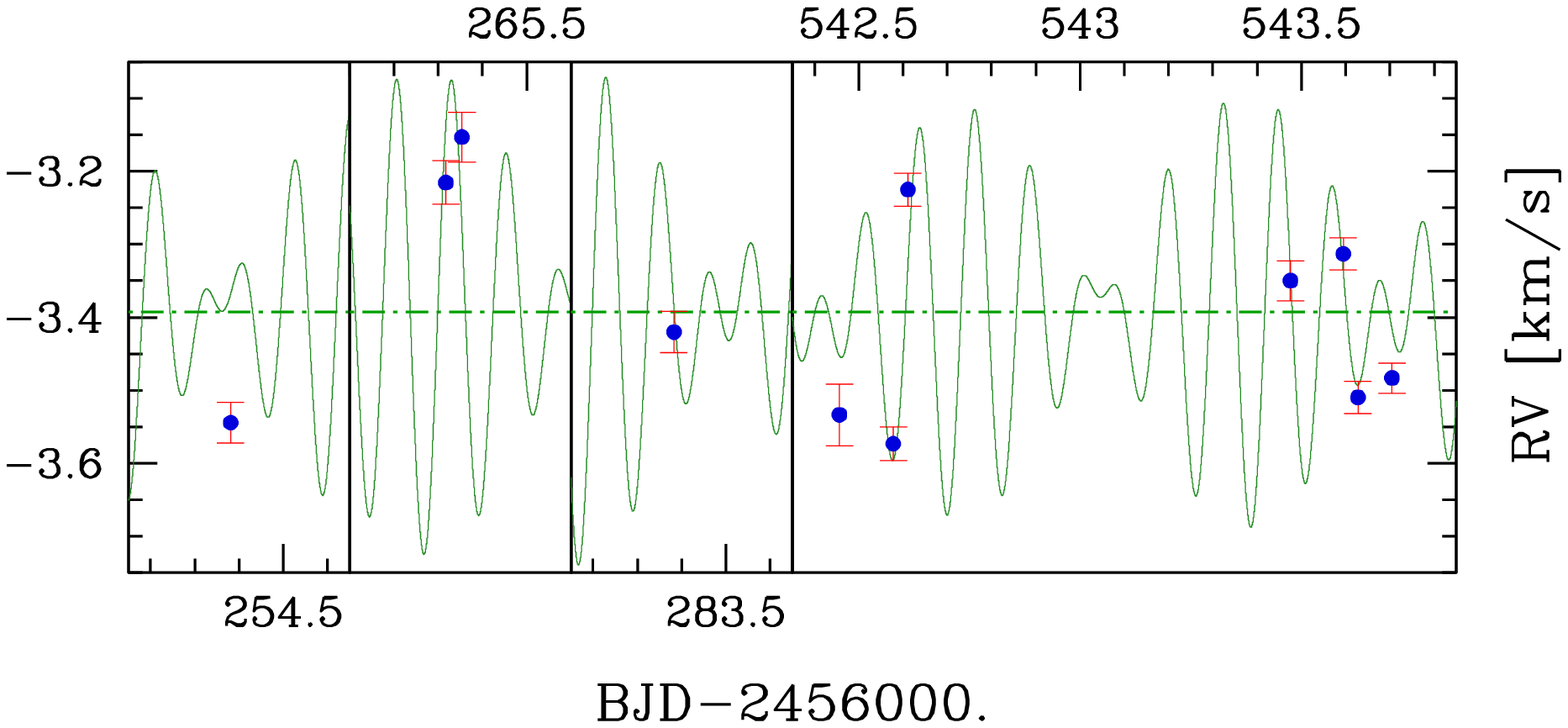}
\centering
\includegraphics[width=0.99\linewidth]{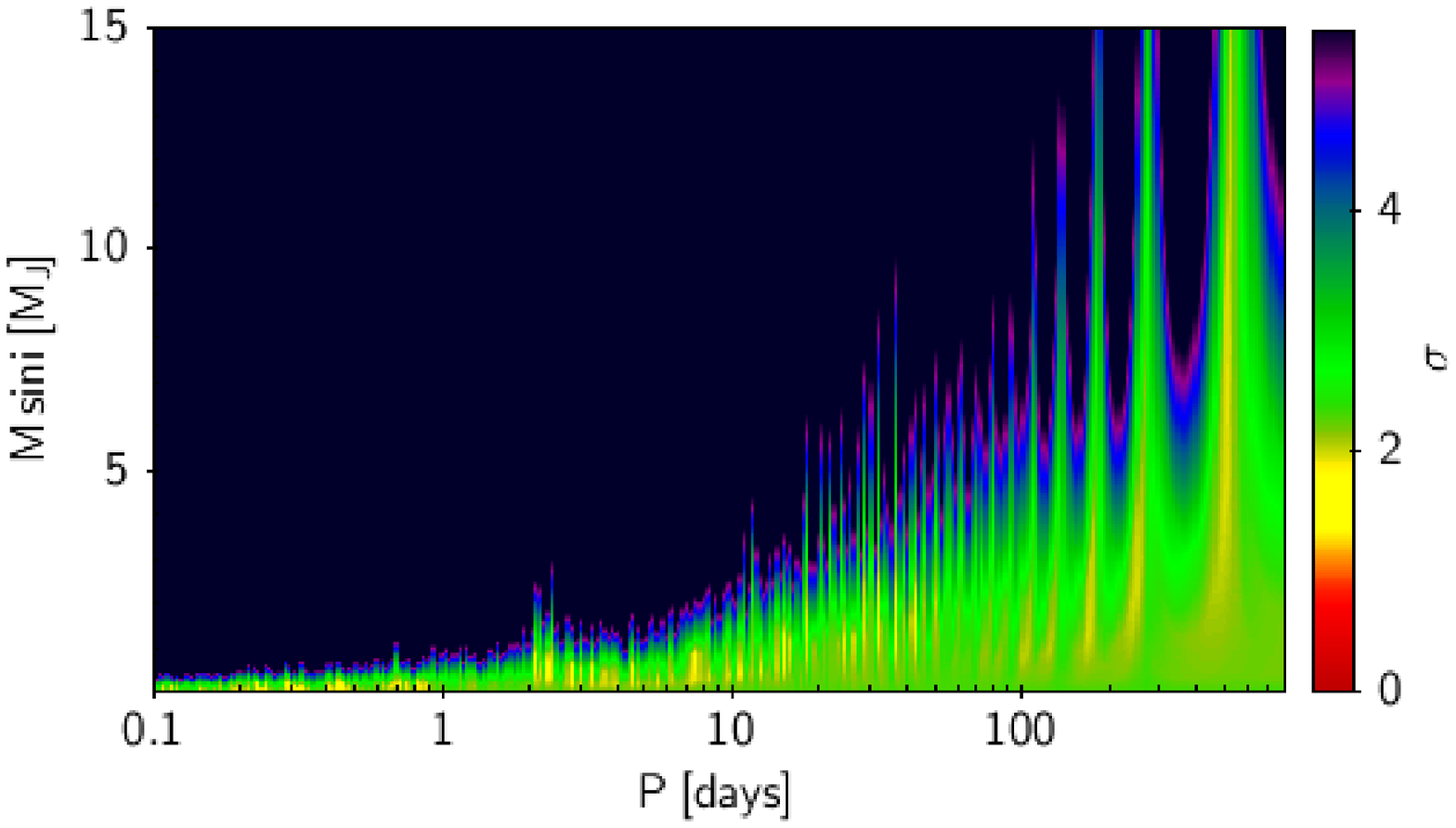}
\caption{HD\,4539. Top: RV measurements and multi-sinusoidal fit corresponding 
to the three main pulsation frequencies (see \citet{silvotti19} for more 
details). The horizontal scale is the same in all epochs. 
Bottom: minimum companion mass vs. orbital period. This plot was obtained 
from the residuals, after having subtracted from the RV measurements the
multi-sinusoidal fit shown in the upper panel.}
\label{fig1}
\end{figure}

\begin{figure}[p]
\centering
\includegraphics[width=1.0\linewidth]{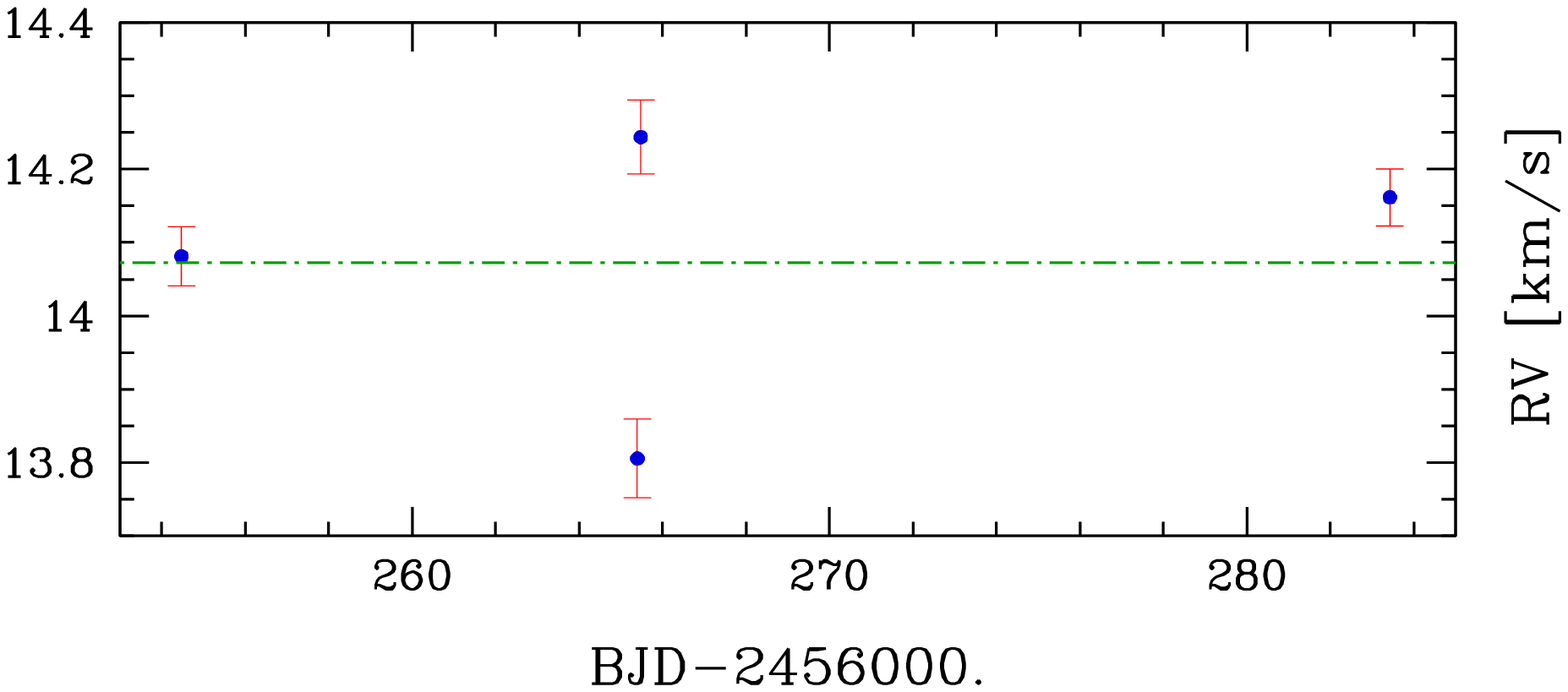}
\includegraphics[width=0.99\linewidth]{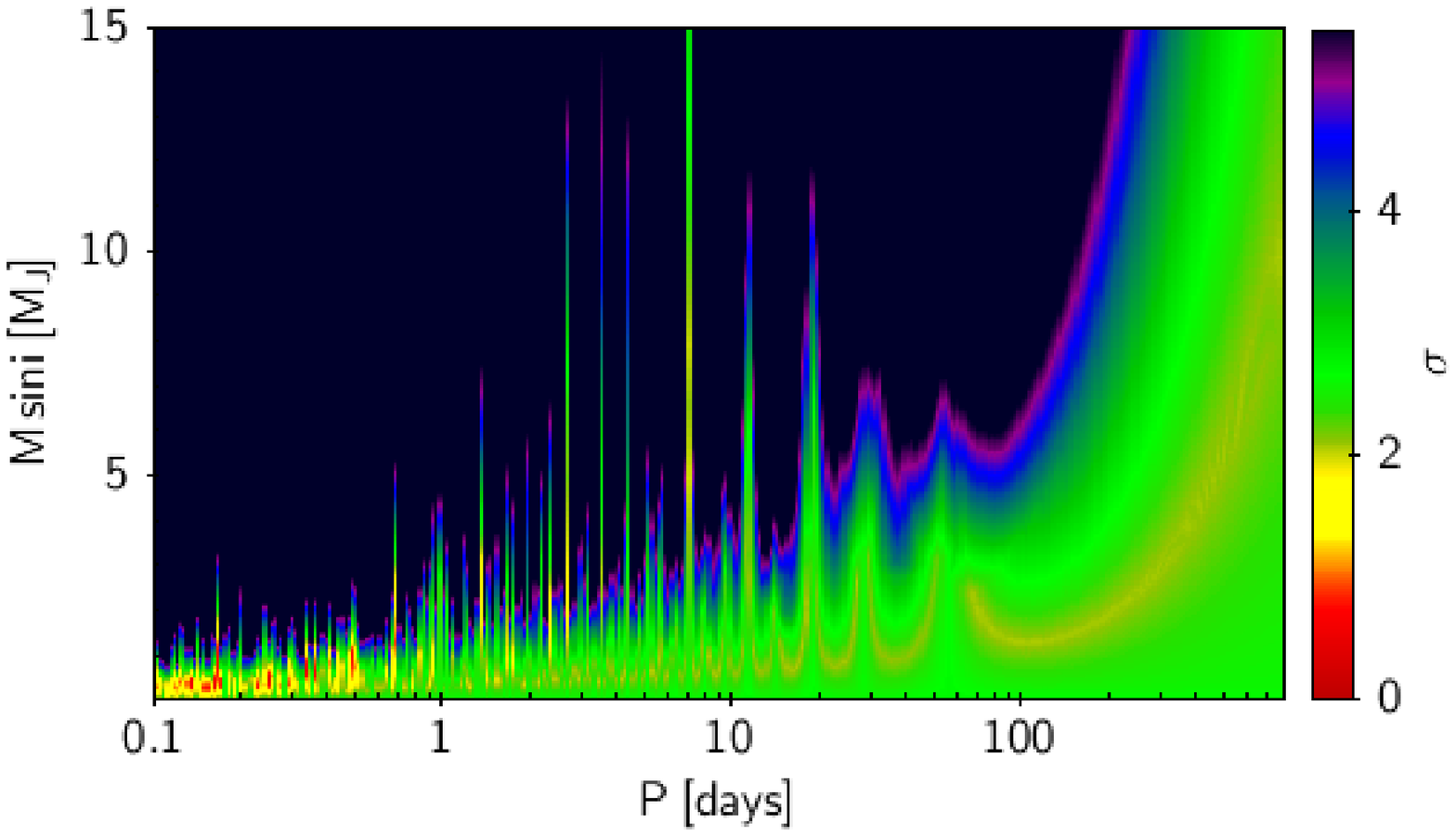}
\caption{PG\,0342+026. Top: RV measurements. 
Bottom: maximum companion mass vs. orbital period.}
\label{fig2}
\end{figure}

\begin{figure}[p]
\vspace{-7mm}
\centering
\includegraphics[width=1.0\linewidth]{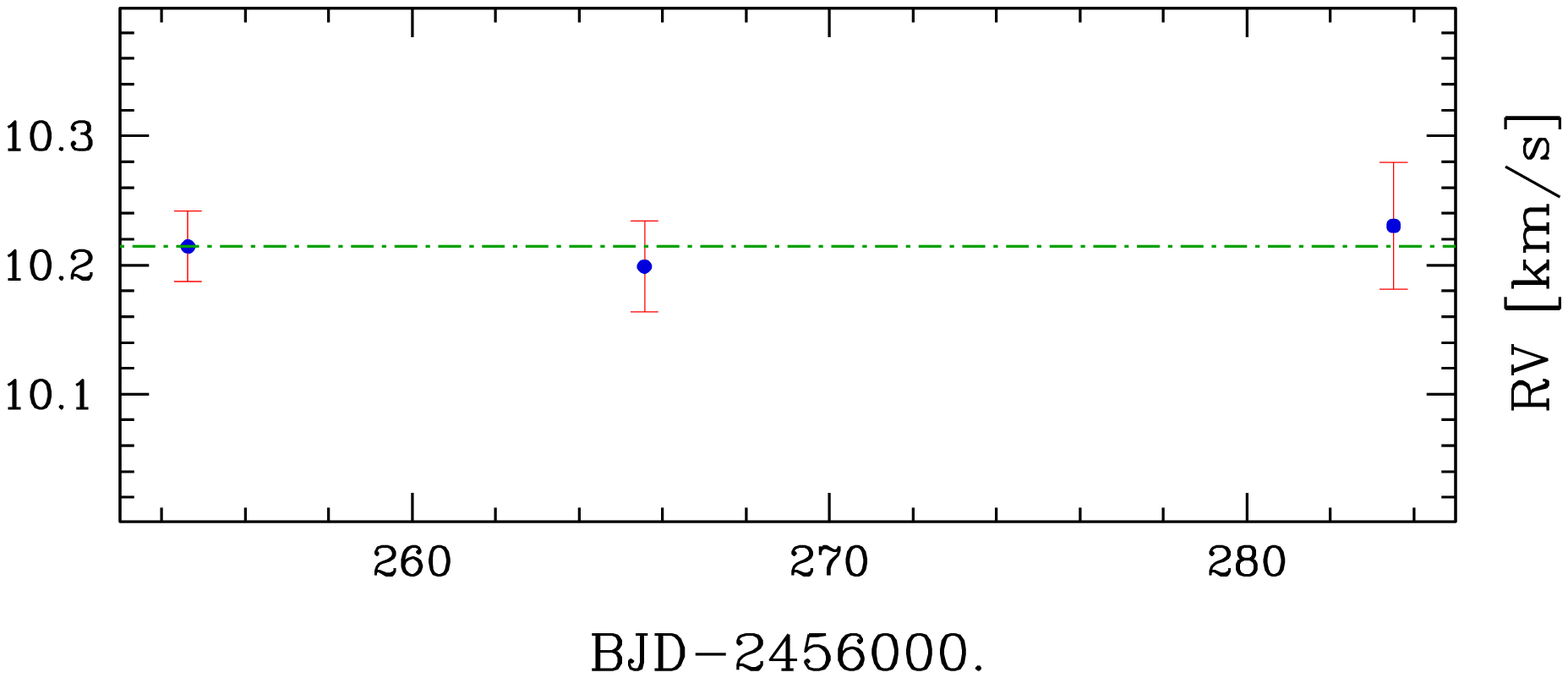}
\includegraphics[width=0.99\linewidth]{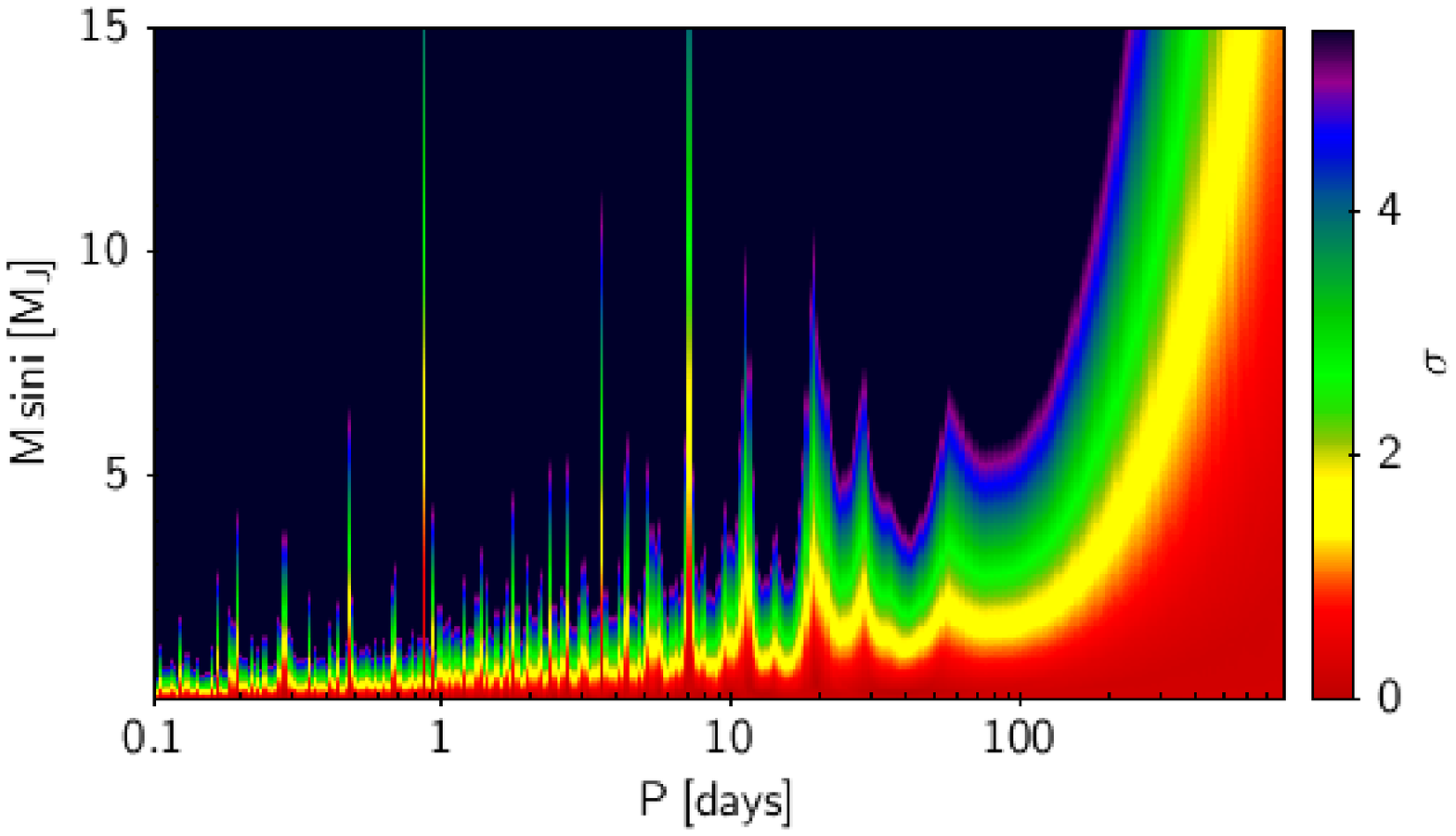}
\caption{Same as Fig.~\ref{fig2} for UVO\,0512-08.}
\vspace{50mm}
\label{fig3}
\end{figure}

\begin{figure}[p]
\vspace{-64mm}
\centering
\includegraphics[width=1.0\linewidth]{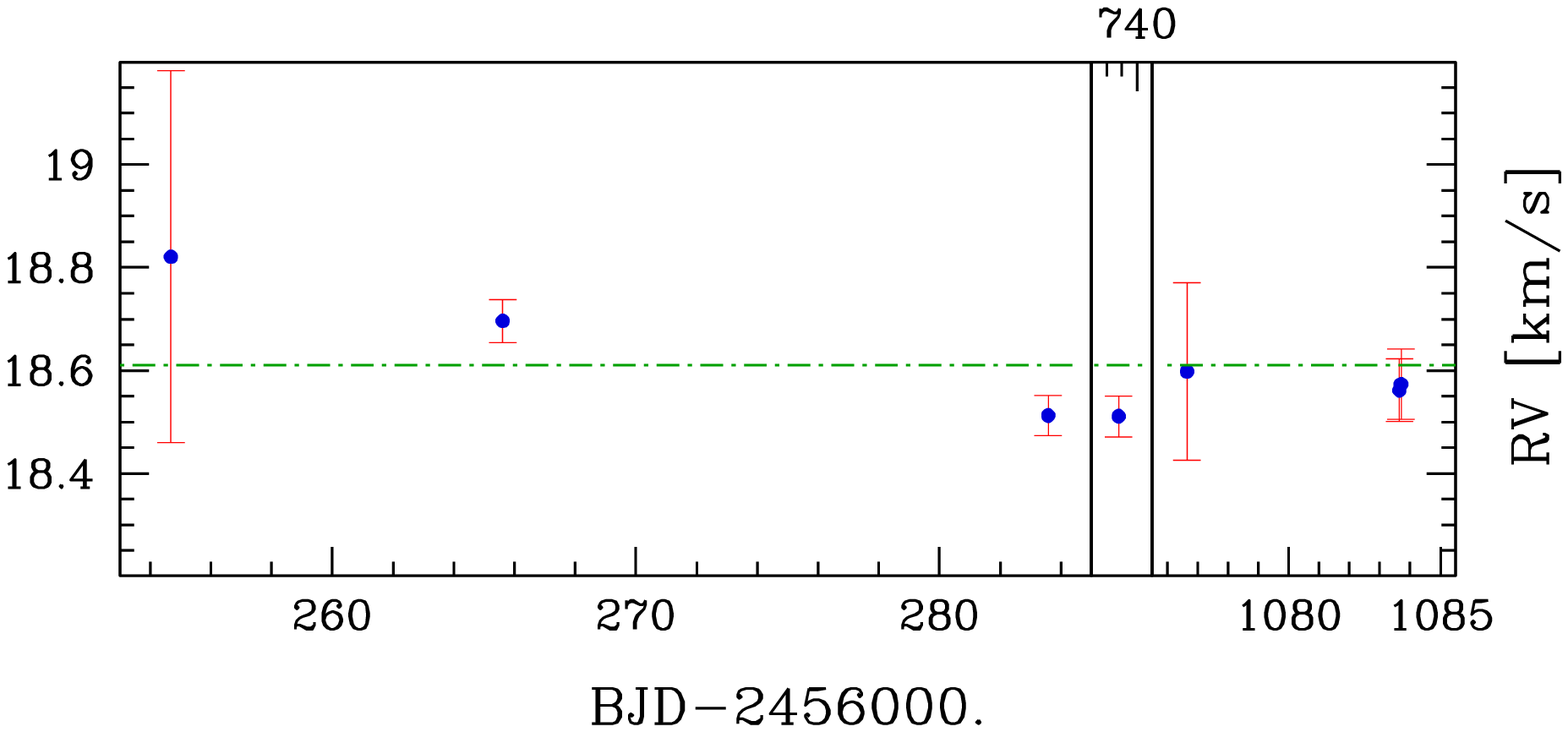}
\includegraphics[width=0.99\linewidth]{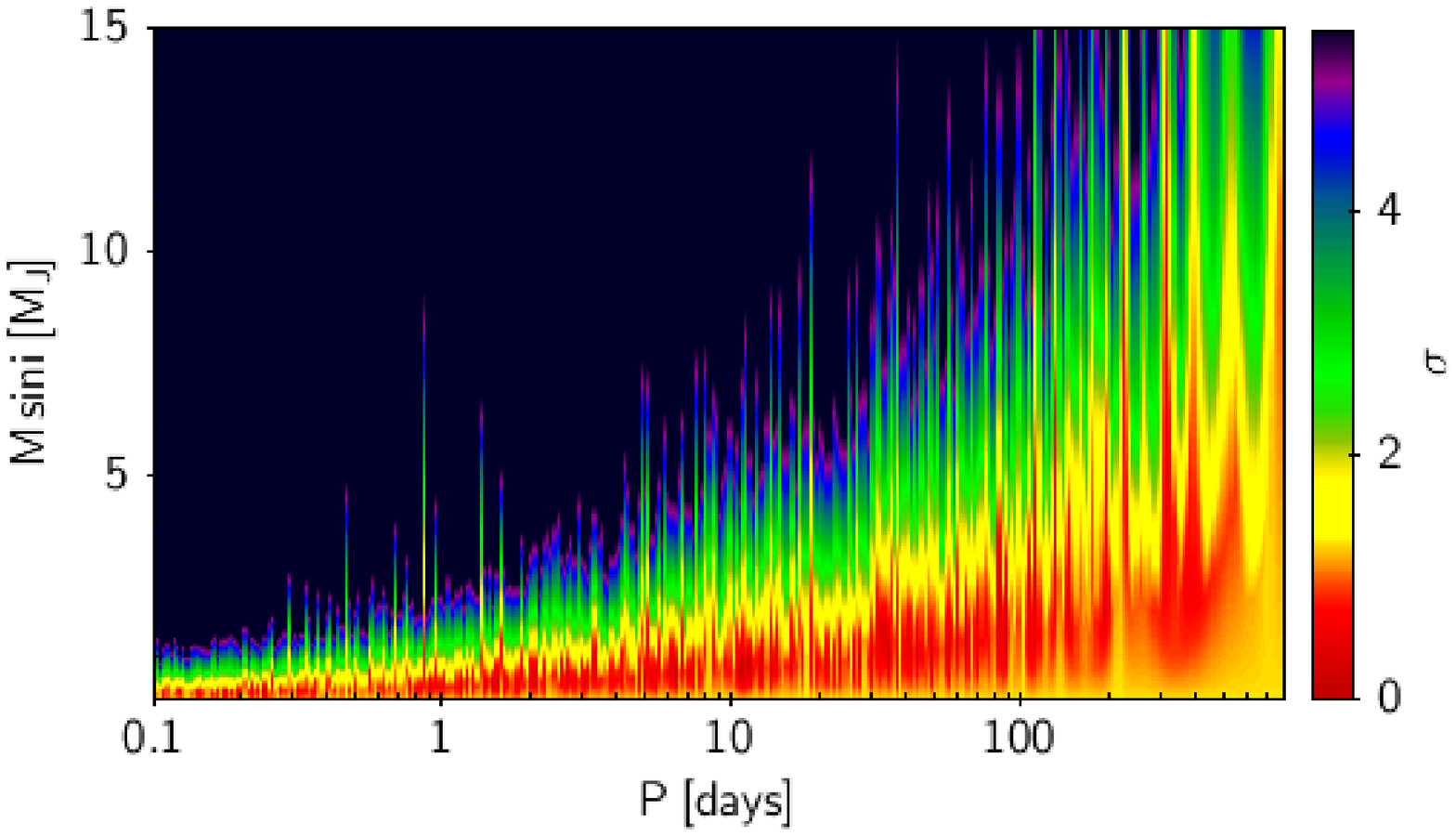}
\caption{Same as Fig.~\ref{fig2} for PG\,0909+276.
The horizontal scale in the upper panel is the same in all epochs.}
\label{fig4}
\end{figure}

\begin{figure}[p]
\centering
\includegraphics[width=1.0\linewidth]{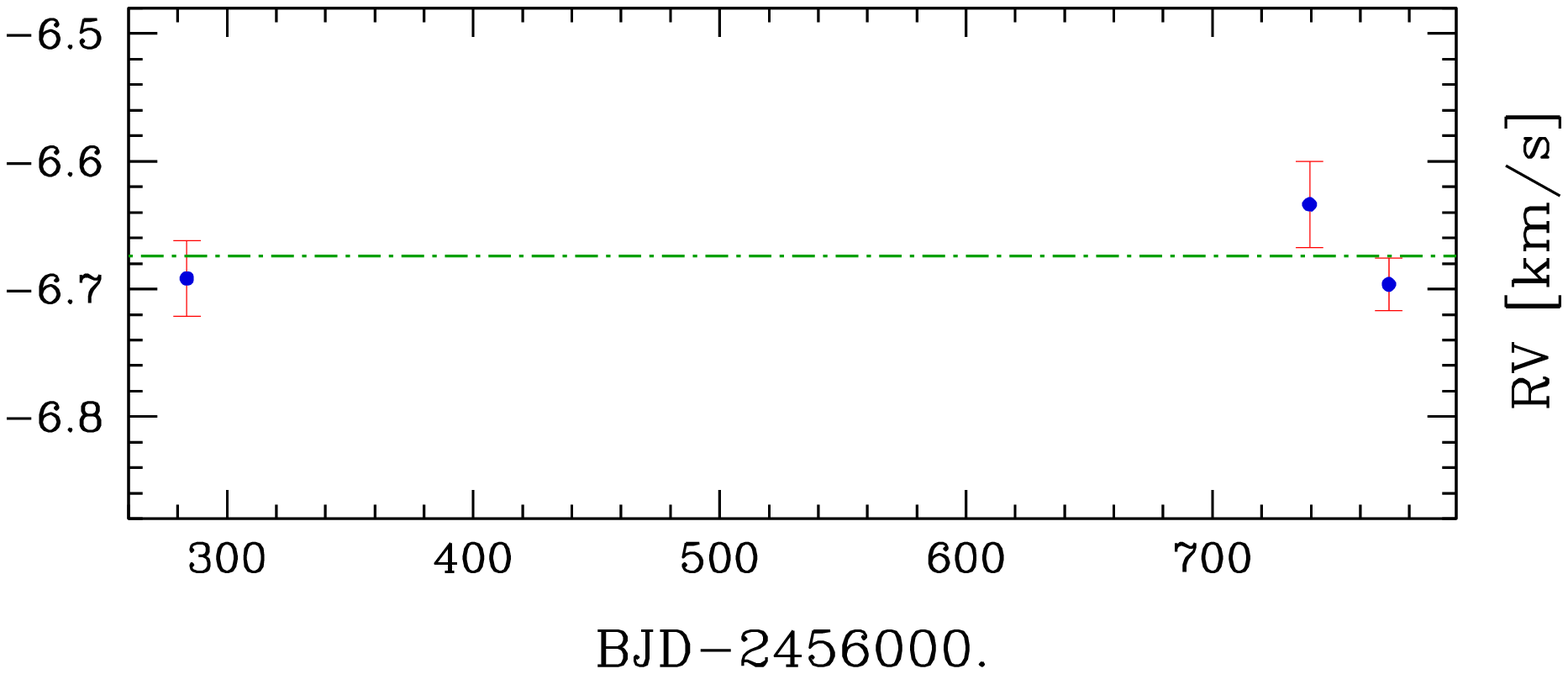}
\includegraphics[width=0.99\linewidth]{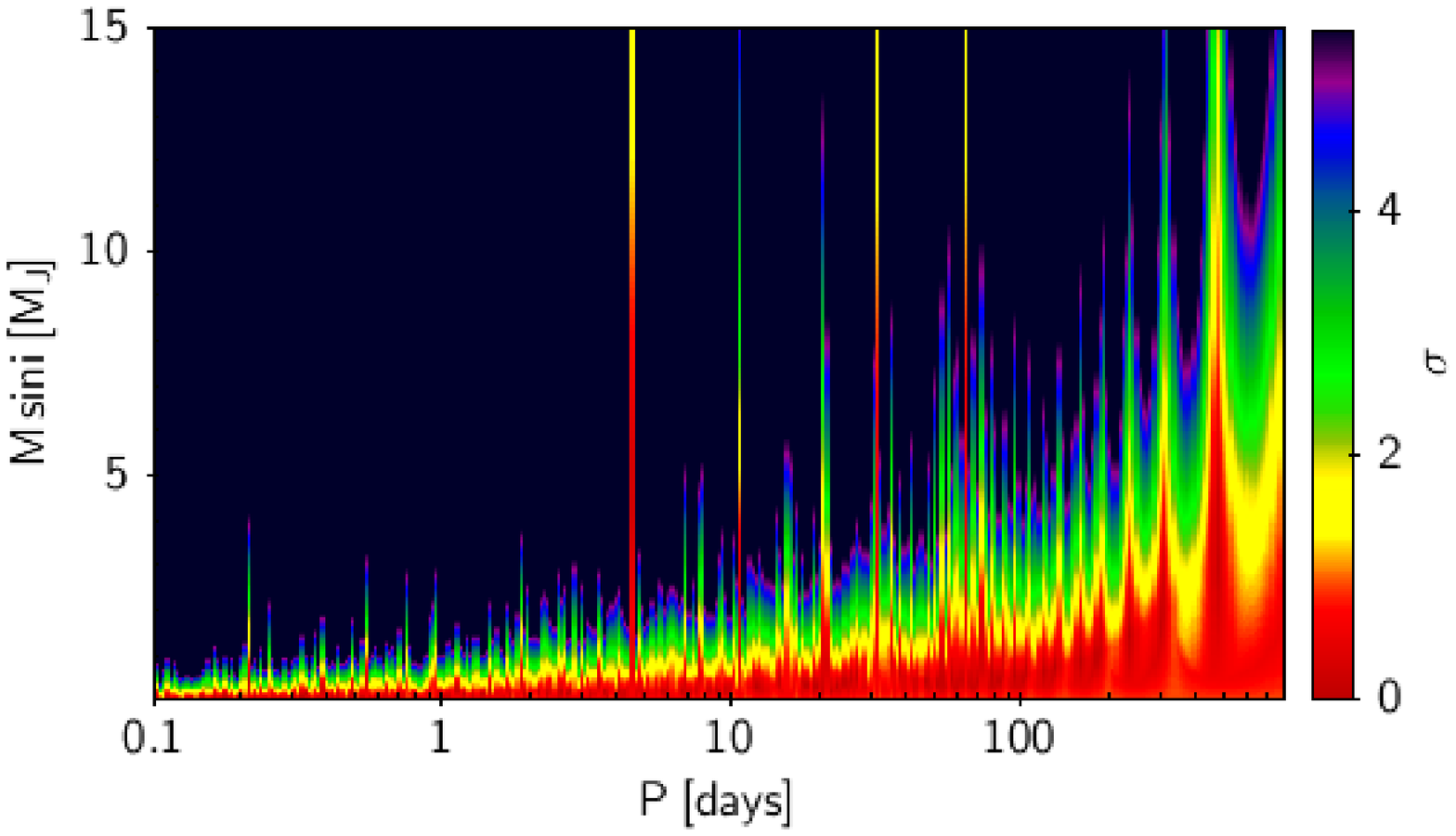}
\caption{Same as Fig.~\ref{fig2} for PG\,1234+253.}
\label{fig5}
\end{figure}

\begin{figure}[p]
\centering
\includegraphics[width=1.0\linewidth]{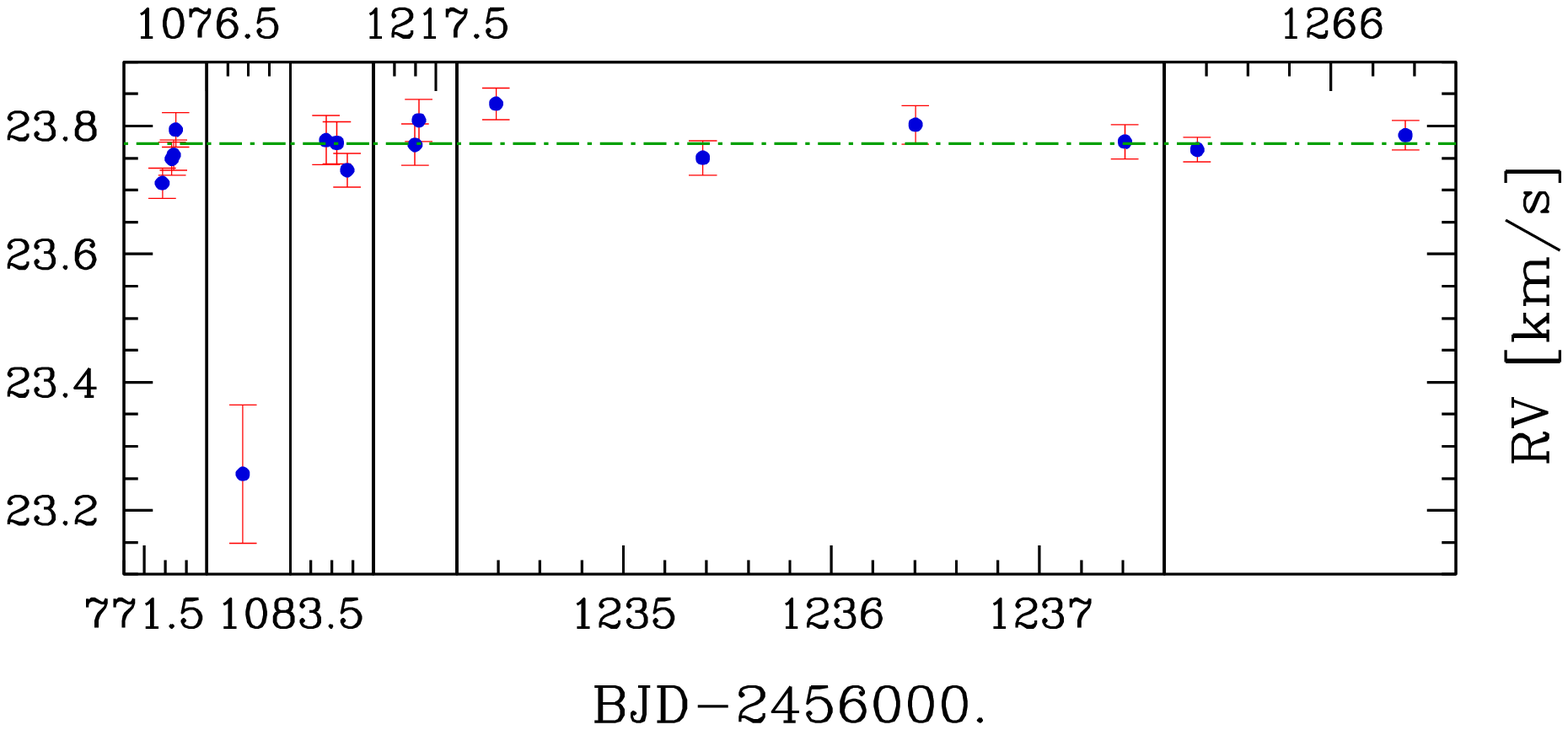}
\includegraphics[width=0.99\linewidth]{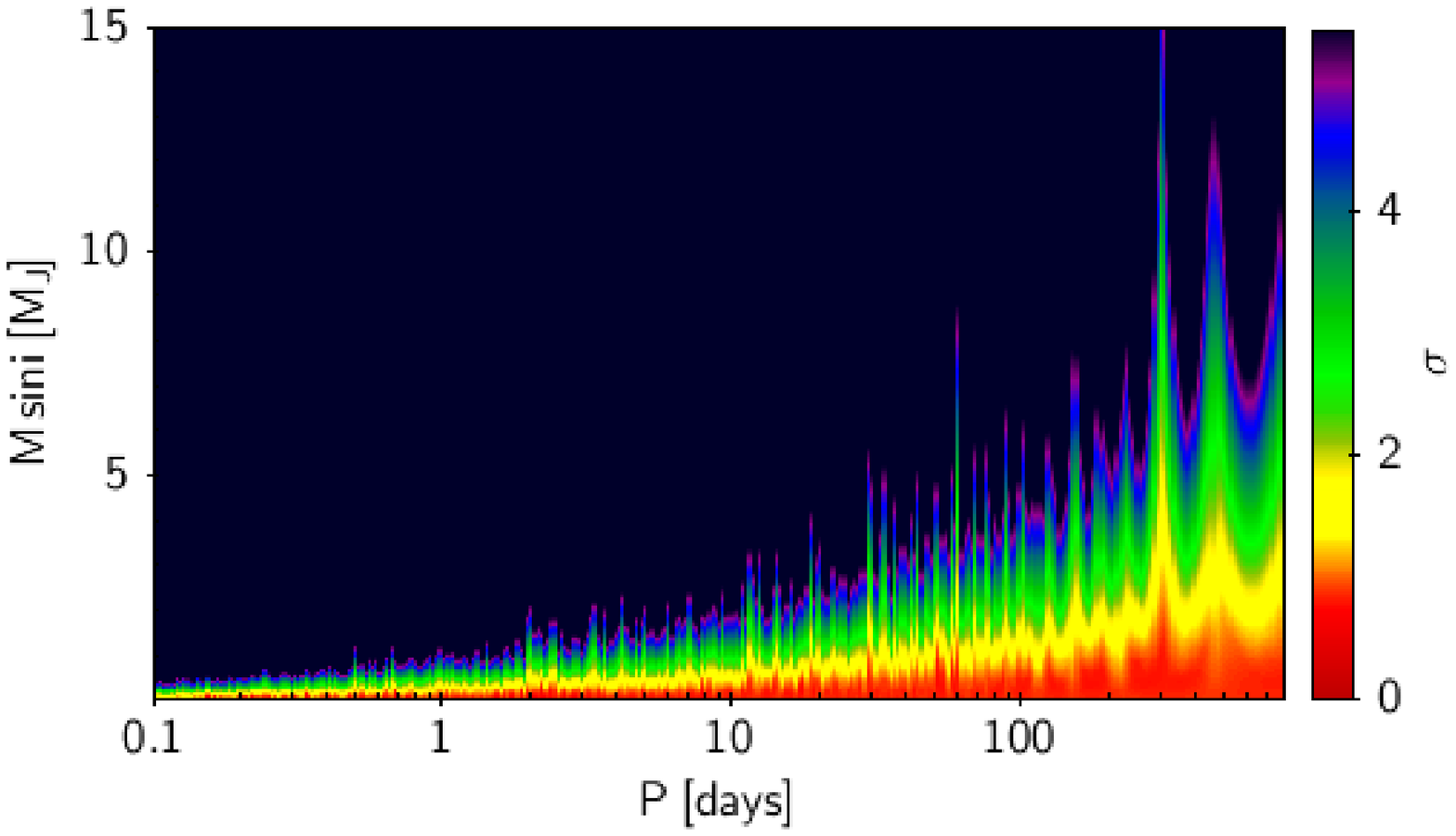}
\caption{Same as Fig.~\ref{fig2} for HD\,149382.
The horizontal scale in the upper panel is the same in all epochs.
The lower plot was obtained excluding the RV outlier at BJD 2457076.67.}
\label{fig6}
\end{figure}

\begin{figure}[p]
\centering
\includegraphics[width=1.0\linewidth]{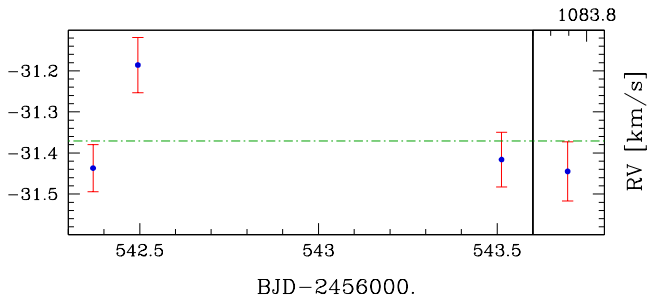}
\includegraphics[width=0.99\linewidth]{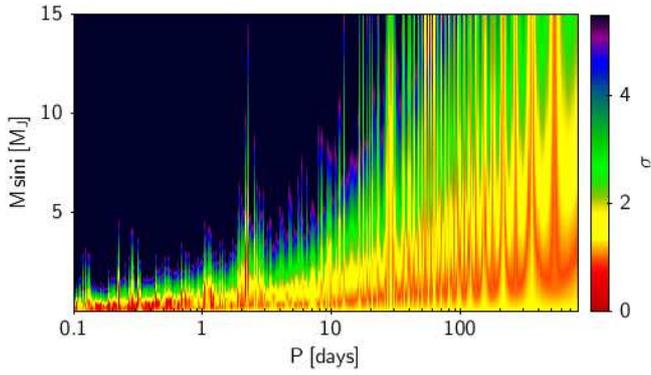}
\caption{Same as Fig.~\ref{fig2} for PG\,1758+364.
The horizontal scale in the upper panel is the same in all epochs.}
\label{fig7}
\end{figure}

\begin{figure}[p]
\centering
\includegraphics[width=1.0\linewidth]{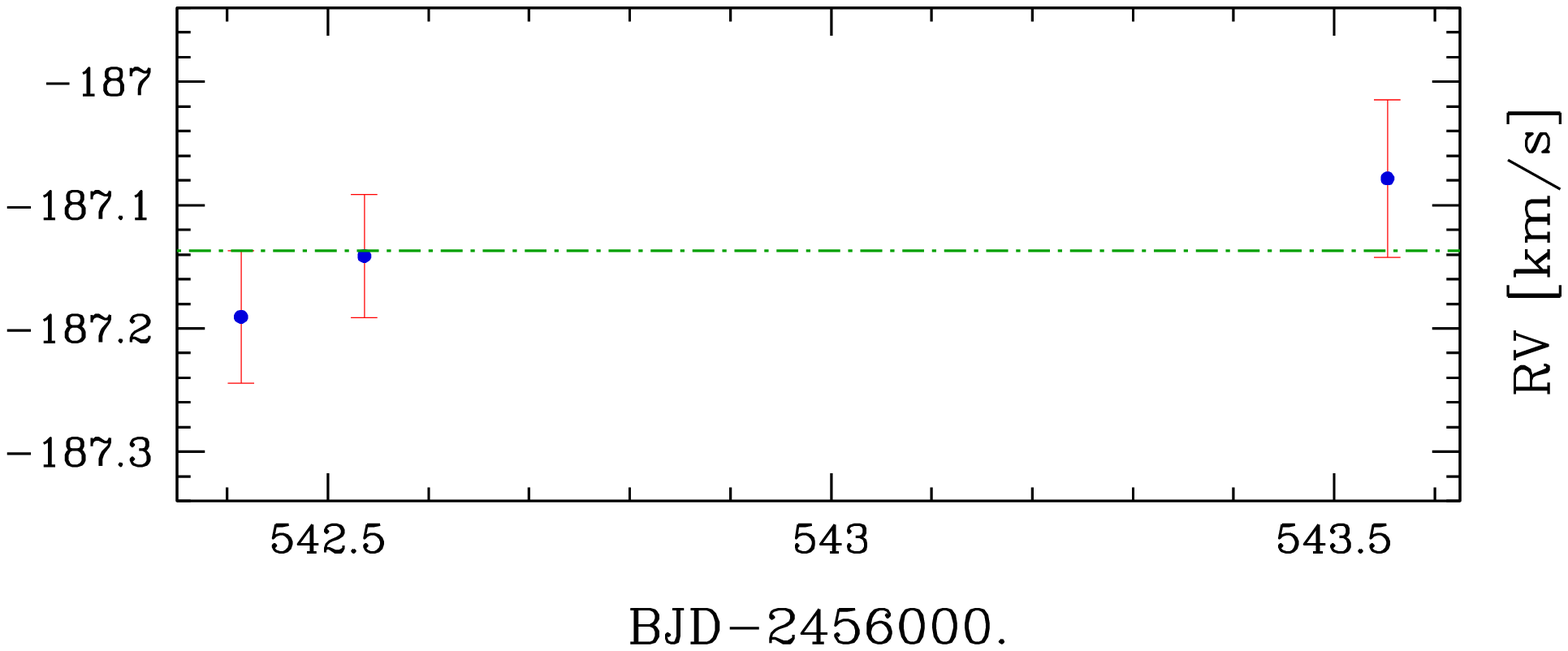}
\includegraphics[width=0.99\linewidth]{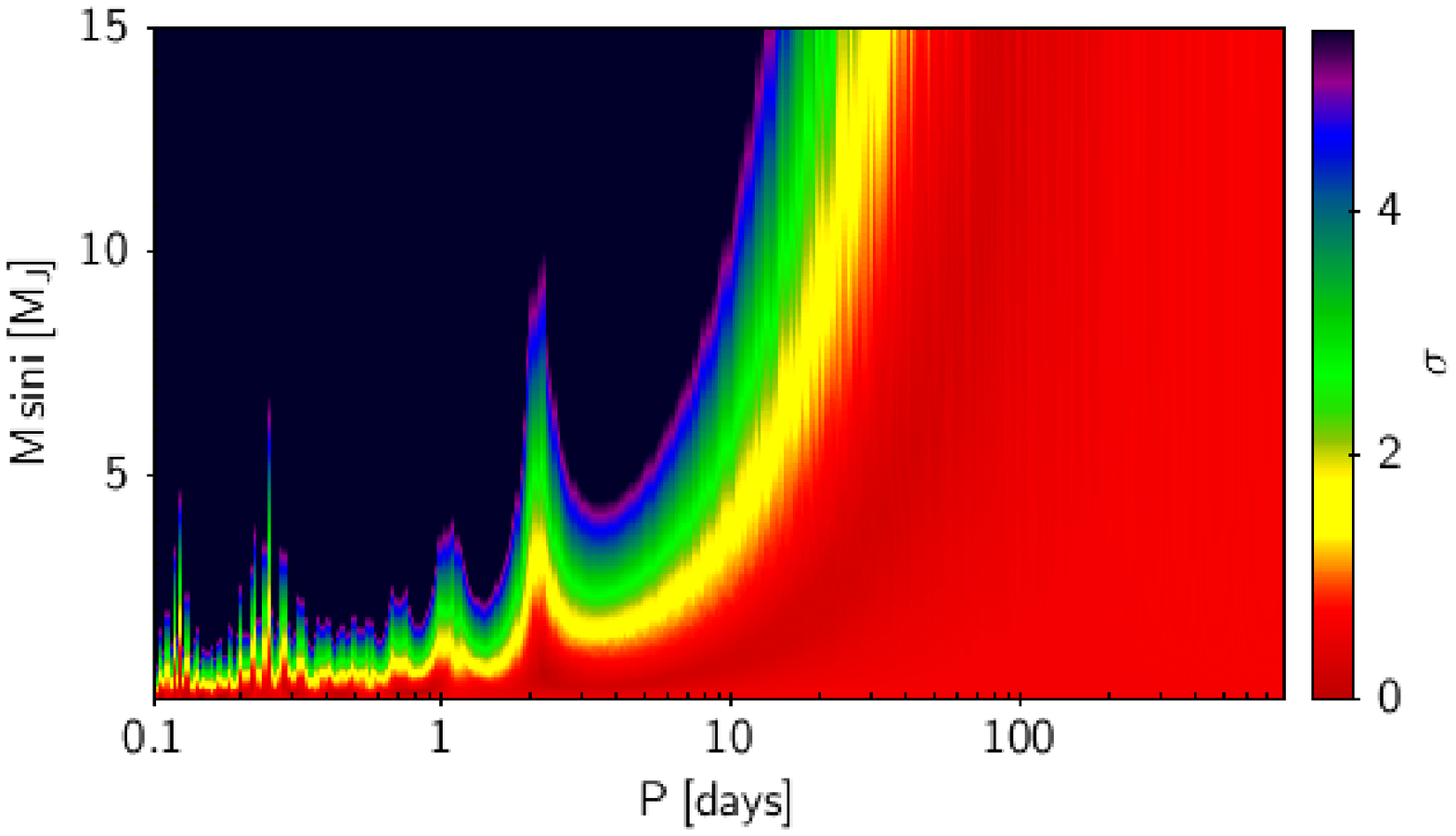}
\caption{Same as Fig.~\ref{fig2} for BD\,48$^\circ$2721.}
\label{fig8}
\end{figure}

For each synthetic RV curve we selected the phase that gives the best fit to 
the data using a weighted least squares algorithm. 
For each observational point we computed the difference, 
in absolute value and in $\sigma$ units (where $\sigma$ is the RV 
error), between the radial velocity and the synthetic RV value. 
The color coding in Fig.~\ref{fig1}--\ref{fig8} (lower panels) corresponds to 
the mean value of this difference in $\sigma$ units.
Note that the minima that we see in these plots do not mean that a real
signal is there but rather that at those orbital periods the sampling of 
the data does not allow us to exclude a companion.
For the two g-mode pulsators, HD\,4539 and PG\,0342+026, we should keep in 
mind that these upper limits to the mass of a companion are likely 
overestimated given that most if not all the variations that we see in 
Fig.~\ref{fig1} and \ref{fig2} are caused by the pulsations.
For HD\,4539, with 11 RV measurements, it was possible to partially subtract
the contribution due to the pulsations (see Fig.~\ref{fig1} and
\citet{silvotti19} for more details).

\begin{figure}[tp]
\centering
\includegraphics[width=0.99\linewidth]{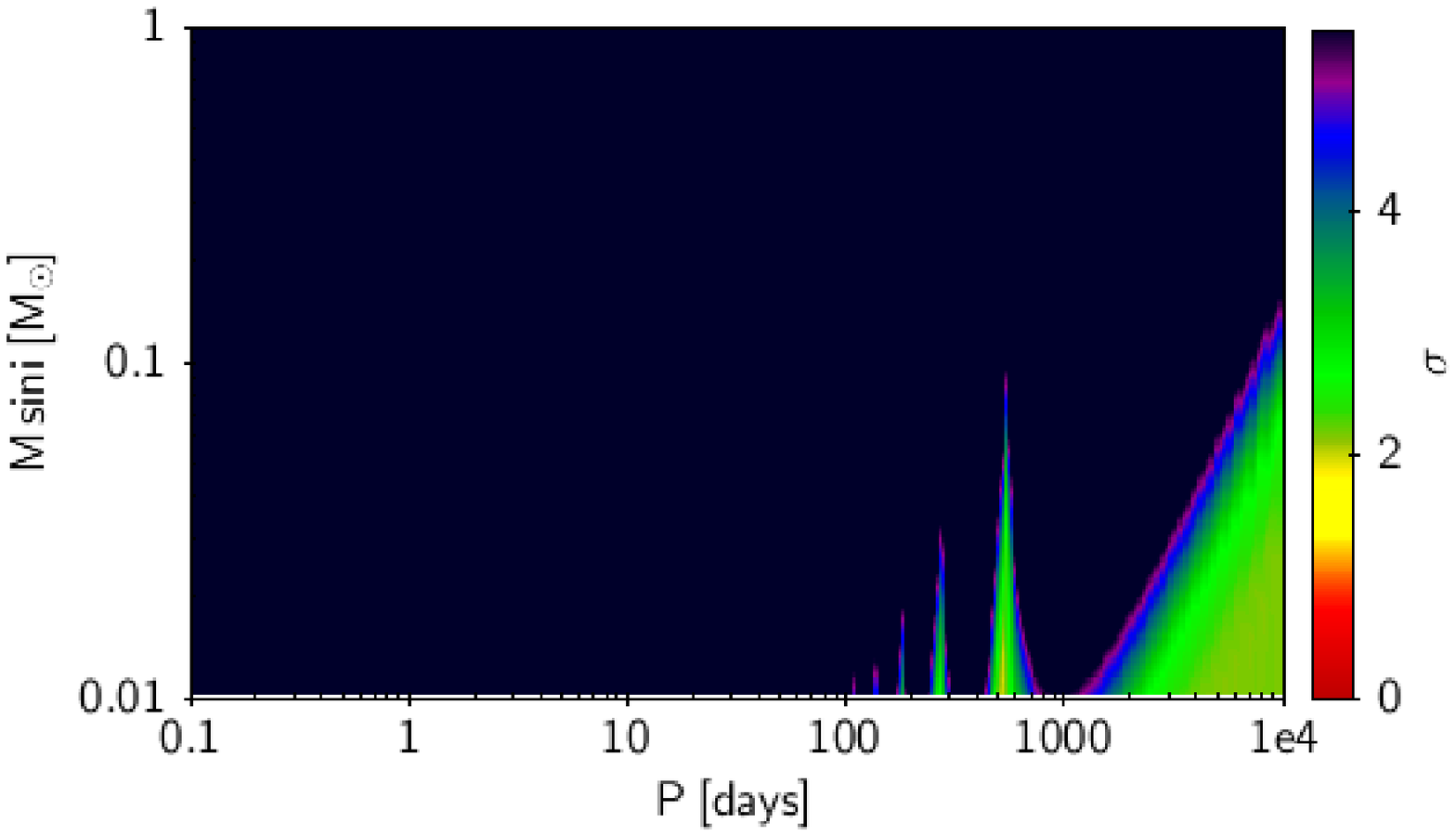}
\caption{HD\,4539: upper limits to the mass of a massive companion in a large 
orbit. Note that in this plot the vertical scale is logarithmic and in solar 
mass units.}
\label{fig9}
\end{figure}

\begin{figure}[p]
\centering
\includegraphics[width=0.99\linewidth]{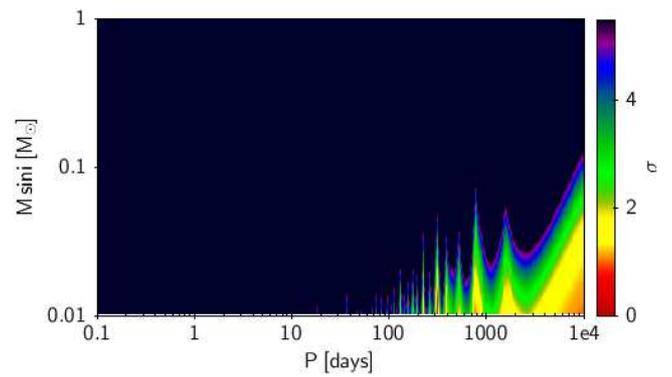}
\caption{Same as Fig.~\ref{fig9} for PG\,0909+276.}
\label{fig10}
\end{figure}

\begin{figure}[p]
\centering
\includegraphics[width=0.99\linewidth]{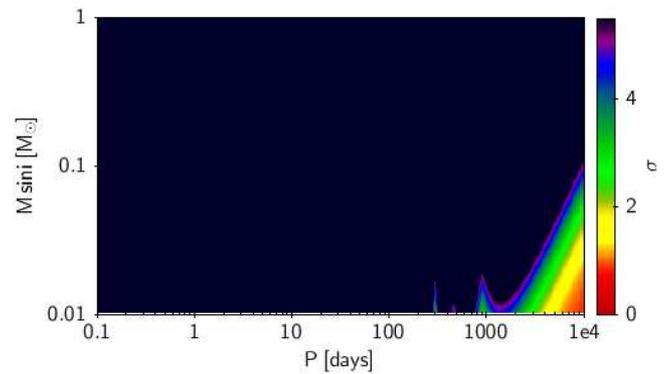}
\caption{Same as Fig.~\ref{fig9} for HD\,149382.}
\label{fig11}
\end{figure}

\begin{figure}[p]
\centering
\includegraphics[width=0.99\linewidth]{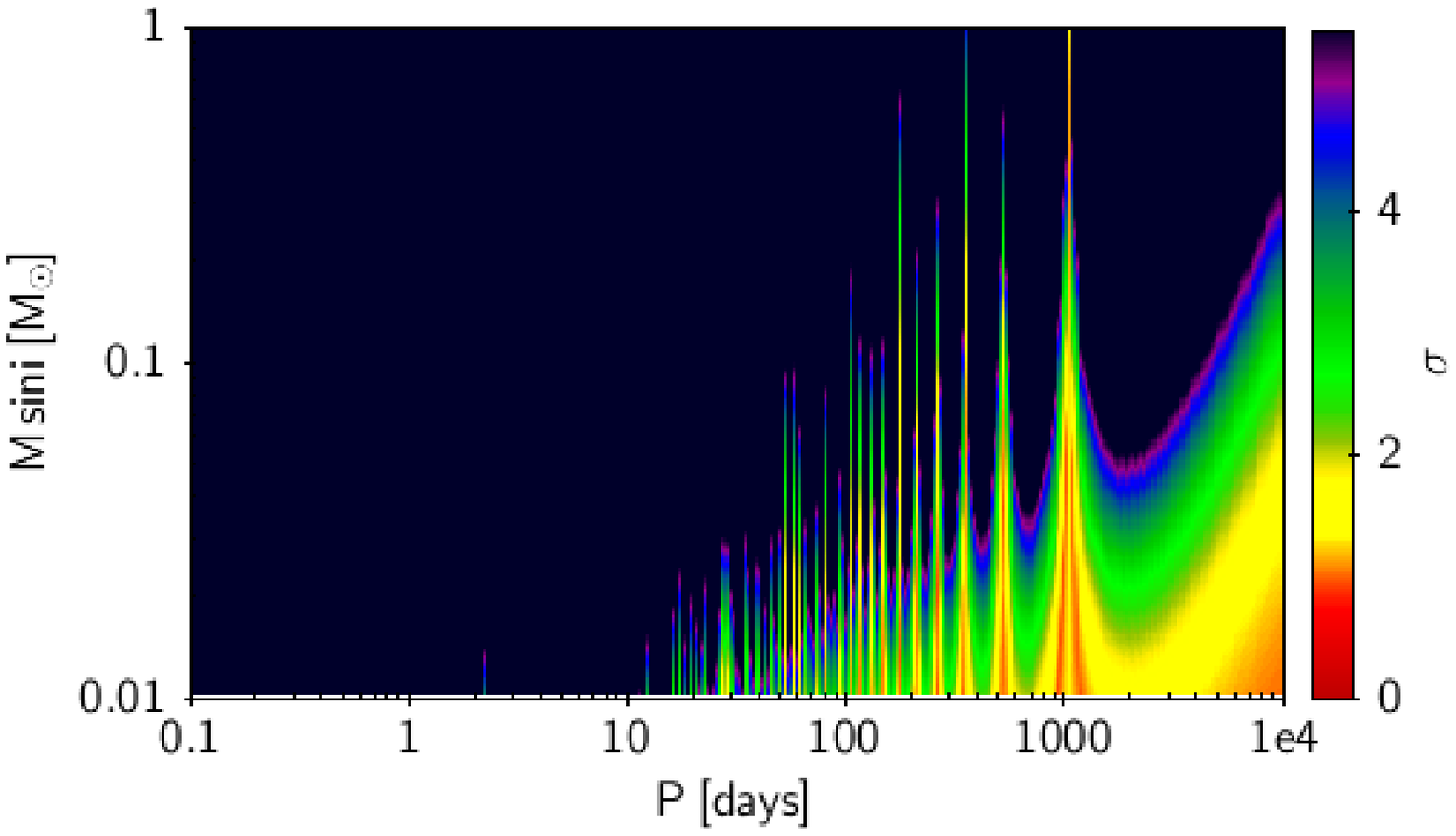}
\caption{Same as Fig.~\ref{fig9} for PG\,1758+364.}
\label{fig12}
\end{figure}

With the same method we computed also the upper limits for a more massive 
unseen companion in a larger orbit but this was possible only for the four 
stars that have a sufficiently long data coverage. We see in 
Fig.~\ref{fig9}--\ref{fig12} that also for these wide binaries we can obtain 
good constraints.

\section{$TESS$ data}

In order to further exclude the presence of a close companion,
it is very usefull to have also high-precision photometric time series,
that can detect transits/eclipses or reflection effects (ellipsoidal 
deformations or Doppler beaming are very weak effects with a low-mass 
companion, see e.g. \citet{esteves15}).
Apart from the two g-mode pulsators observed by $K2$ or $TESS$, another
star, UVO0512-08, was observed by the $TESS$ mission in sector 5 (see next
subsection), while all the other five stars will be observed by $TESS$ in the 
next months (sectors 21-26).

\subsection{$TESS$ observation of UVO0512-08}

\begin{figure}[h]
\centering
\includegraphics[width=1.0\linewidth]{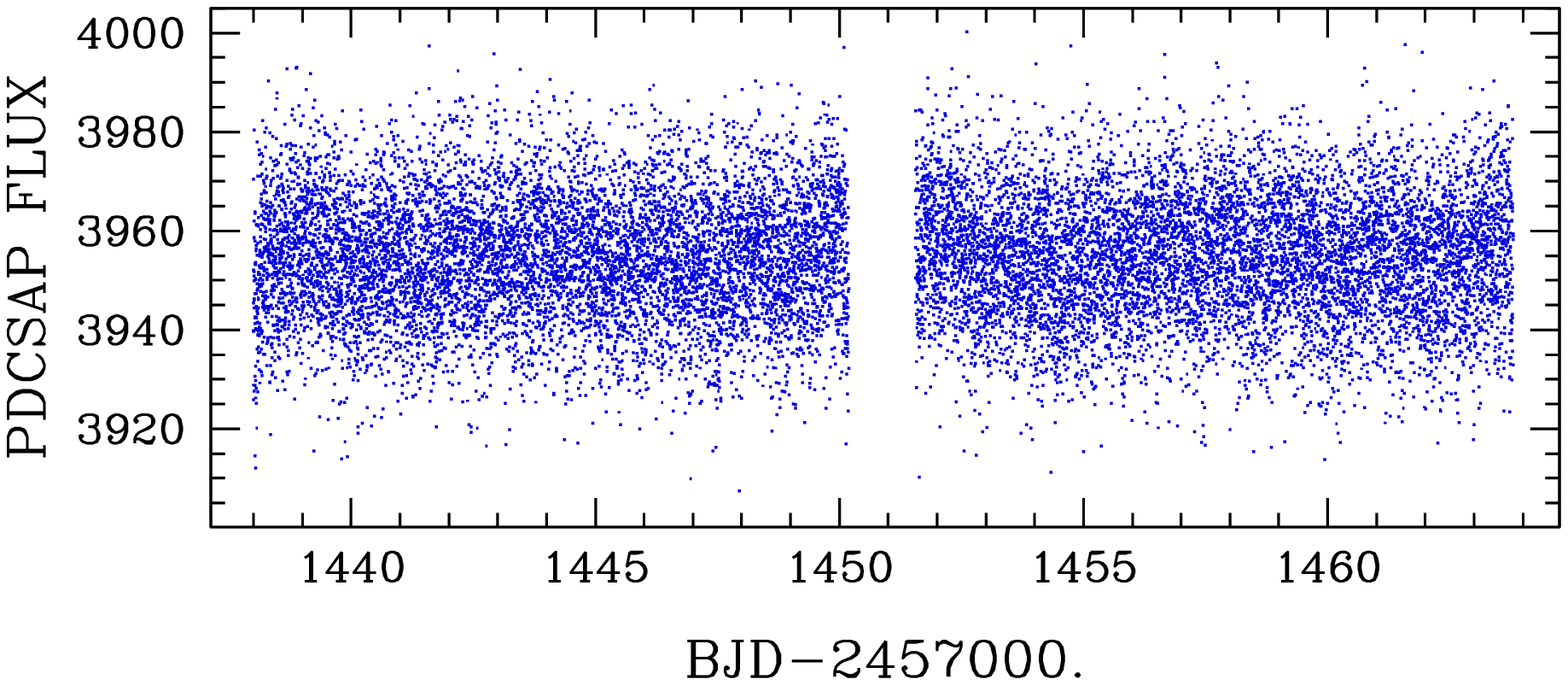}
\includegraphics[width=1.0\linewidth]{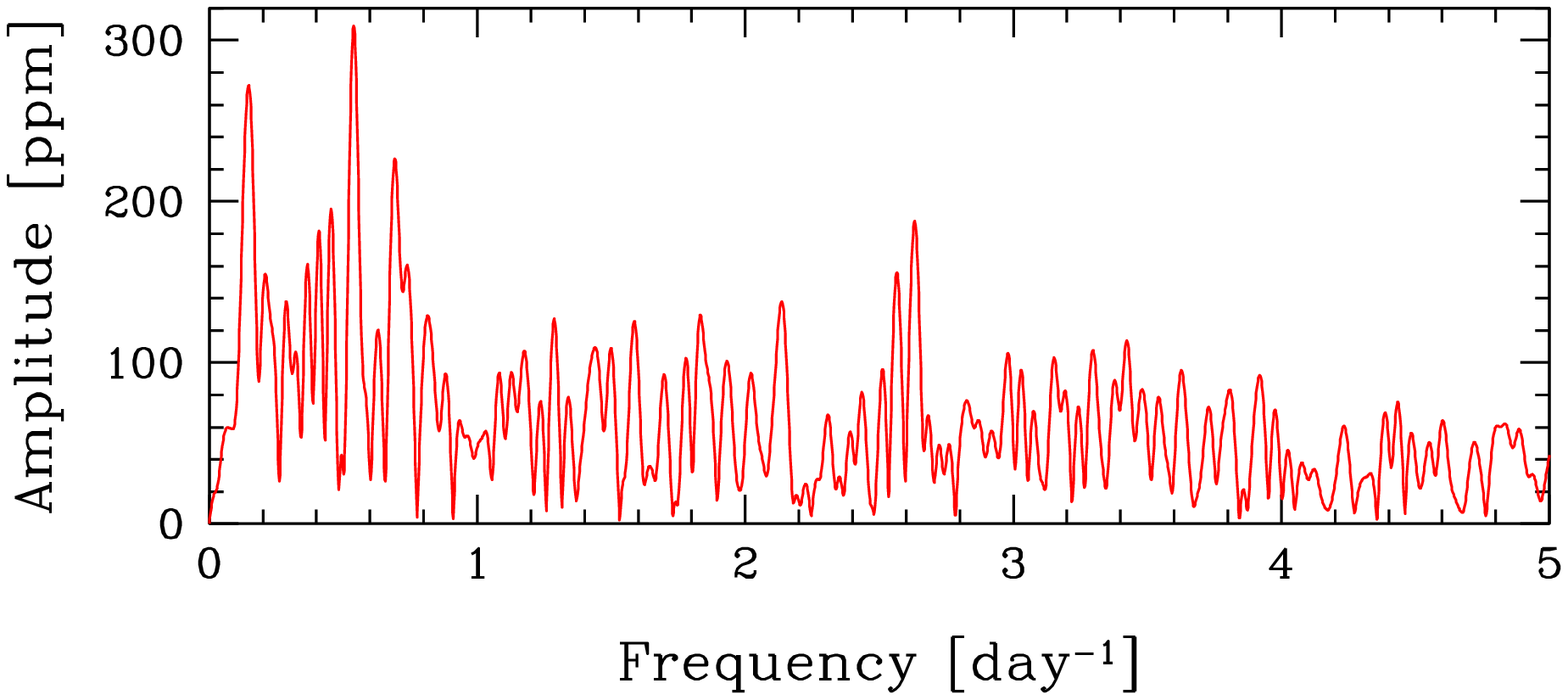}
\caption{Top: $TESS$ light curve of UVO0512-08.
Bottom: amplitude spectrum.}
\label{fig13}
\end{figure}

The $TESS$ data on UVO0512-08 are shown in the upper panel of Fig.~\ref{fig13}.
We used the PDCSAP (Pre-search Data Conditioning Single Aperture) flux 
removing a few outlayers and removing also the very last part of the run which 
is noisy.
The amplitude spectrum shows only a few peaks at low frequency
(lower panel of Fig.~\ref{fig13}) and allows us to exclude at least one of the 
two main peaks that we see in the lower panel of Fig.~\ref{fig3}, the one with 
a period of about 0.87~d.
The other peak near 7.3~d, however, is partially compatible with a 
low-amplitude peak in Fig.~\ref{fig13} at $\sim$0.15~d$^{-1}$.
On the other hand, the main peak in the amplitude spectrum of Fig.~\ref{fig13},
at about 0.54~d$^{-1}$, does not have any counterpart in the lower panel of 
Fig.~\ref{fig3}.

\section{Conclusions}

High resolution spectroscopy is a good way to set strong limits to
the mass of a companion to sdB stars.
To have also high-quality light curves such as those provided by $Kepler$/$K2$
or $TESS$ make these limits even stronger.
In particular, with $TESS$, this will be possible for almost all the bright
sdB stars in both emispheres.
The present limits keep open the unresolved question of the mechanism 
that caused the loss of the envelope for the giant precursors of these stars.
To obtain further data on these height stars and to extend this programme
to a larger sample of bright, apparently single, sdB stars would help
to make progress in understanding the evolution of these stars.








\section*{Acknowledgments}
We thank the TNG staff for the great technical support, the TASOC WG8 people
for the effort in proposing and organizing the TESS observations of many sdB 
stars including UVO0512-08, and St\'ephane Charpinet and collaborators for 
organizing a wonderful sdOB meeting at Hendaye in June 2019.


\end{document}